\documentclass[aip,graphicx]{revtex4-1}
\usepackage[cp1251]{inputenc}
\usepackage[T2A]{fontenc}
\usepackage[english]{babel}
\usepackage{amssymb,latexsym,amsmath,amscd}
\usepackage{graphicx,color,framed}
\begin{document}
\selectlanguage{english}
\title{Nonlocal statistical field theory of dipolar particles in electrolyte solutions}
\author{\firstname{Yury} \surname{Budkov}}
\email[]{ybudkov@hse.ru}
\affiliation{School of Applied Mathematics, Tikhonov Moscow Institute of Electronics and Mathematics, National Research University Higher School of Economics, Tallinskaya st. 34, 123458 Moscow, Russia}
\affiliation{G.A. Krestov Institute of Solution Chemistry of the Russian Academy of Sciences, Akademicheskaya st. 1, 153045 Ivanovo, Russia}
\begin{abstract}
We present a nonlocal statistical field theory of a dilute electrolyte solution with small additive of dipolar particles. We postulate that every dipolar particle is associated with an arbitrary probability distribution function (PDF) of distance between its charge centers. Using the standard Hubbard-Stratonovich transformation, we represent the configuration integral of the system in the functional integral form. We show that in the limit of a small permanent dipole moment, the functional in integrand exponent takes the well known form of the Poisson-Boltzmann-Langevin (PBL) functional. In the mean-field approximation we obtain a non-linear integro-differential equation with respect to the mean-field electrostatic potential, generalizing the PBL equation for the point-like dipoles obtained first by Abrashkin et al. We apply the obtained equation in its linearized form to derivation of the expressions for the mean-field electrostatic potential of the point-like test ion and its solvation free energy in salt-free solution, as well as in solution with salt ions. For the 'Yukawa'-type PDF we obtain analytic relations for both the electrostatic potential and the solvation free energy of the point-like test ion. We obtain a general expression for the bulk electrostatic free energy of the solution within the Random phase approximation (RPA). For the salt-free solution of the dipolar particles for the Yukawa-type PDF we obtain an analytic relation for the electrostatic free energy, resulting in two limiting regimes. Namely, when the Debye screening length attributed to the charged groups of the dipolar particles is much bigger than the mean-square dipole length, the electrostatic correlations of the dipolar particles consist in their effective Keesom attraction, so that the electrostatic free energy is proportional to the square of dipolar particles concentration. In the opposite case, when the mean-square dipole length is much bigger than the Debye length, the charge centers of dipolar particles can be considered as unbonded ions, so that the electrostatic free energy can be described by the Debye-Hueckel limiting law. Finally, we analyze the limiting laws, following from the general relation for the electrostatic free energy of solution in presence of both the ions and the dipolar particles for the case of Yukawa-type PDF.
\end{abstract}
\maketitle
\section{Introduction}
Development of the polar fluids statistical theory has a long history starting from the classical works of L. Onsager \cite{Onsager1936} and J. Kirkwood \cite{Kirkwood1939}. So far, some progress has been achieved in statistical mechanics of dipolar fluids allowing us to calculate statistical dielectric permittivity of molecular fluids, such as water, aliphatic alcohols, {\sl{etc}} \cite{Onsager1936,Kirkwood1939,Hoye1976,Chandler1977,Holowko_book} with the accuracy appropriate for experimentalists. There are also a small number of theoretical approaches allowing one to calculate self-consistently the electrostatic free energy of a polar fluid alongside the dielectric permittivity. Levin \cite{Levin1999} formulated a theoretical model that allowed him to obtain an analytic expression for the free energy of dipolar hard spheres fluid. Using the idea of Debye charging process \cite{Fisher1993}, he calculated a contribution of a dipole-dipole electrostatic interaction to the total free energy of the dipolar hard spheres based on the Onsager's reaction field theory \cite{Onsager1936}. It should be noted that this approach was earlier applied to the evaluation of the electrostatic free energy for the dipolar hard spheres in work \cite{Nienhuis1972}. It is also important to mention the approach, based on the linearized Poisson-Boltzmann (PB) equation application for the mixture of dipoles with ionic components \cite{Weiss1998}. Within that approach, using the linearized PB equation with undefined constant dielectric permittivity, the electrostatic potentials of the point-like charge and point-like dipole were calculated. Further, based on the Debye-Hueckel approximation, the energies of the electrostatic interaction between the point ion and surrounding ions and dipoles were calculated. The energies of electrostatic interaction between the point dipole and the surrounding dipoles and ions were obtained in the same manner. Further, equating the total interaction energies "ion-dipole" and "dipole-ion", the authors obtained an equation for dielectric permittivity as a function of dipole moment, temperature, dipolar particles concentration and ionic strength, that is a generalization of the classic Onsager's equation \cite{Onsager1936}. The derived equation allowed the authors to calculate the electrostatic contribution to the total free energy and analyse the phase behavior of the electrolyte solution accounting for the formation of ion pairs modelled by dipoles.

There are also a few works, which investigate the mixture of dipolar particles with ions of salt in the confinement or near the charged electrode within the field-theoretical approach. In work \cite{Abrashkin2007} Abrashkin et al. formulated a statistical Poisson-Boltzmann-Langevin (PBL) field theory taking into account explicitly the molecules of polar fluid. In the mean-field approximation the authors derived a modified Poisson-Boltzmann (PB) equation taking into account the dipolar particles, making the local dielectric permittivity of solution dependent on the local electric field. Applying the obtained equation to the description of the electrolyte solution confined between the oppositely charged flat infinite walls, the authors showed that the osmotic pressure calculated within their PBL theory deviated significantly from the value that was obtained in the classical PB theory with a continuous dielectric solvent for rather small distances (a few nanometers) between the charged walls. It should be noted that a similar statistical field theory was earlier obtained by Coalson et al. \cite{Coalson1996} generalizing the classical PB theory and taking into account explicitly the particles with arbitrary electric multipole moments. As an application of their theory, the authors made a numerical calculation of the profiles of the solution local dielectric permittivity between the two flat oppositely charged walls bounding the mixture of ions and dipolar particles, as well as the surface free energy as a function of the distance between the walls. Budkov et al. \cite{Budkov2015} generalized the PBL theory, deriving a modified PB equation with an explicit account of the polarizability tensor of the solvent molecules in addition to their permanent dipole moment within a field-theoretical approach. The authors used their modified PB equation to generalize the classical Gouy-Chapman theory of the electric double layer in the electrolyte solutions for the case when there is a small amount of polarizable/polar co-solvent molecules in the solution. They investigated the influence of the permanent dipole moment and polarizability of the co-solvent molecules on the differential capacitance of the electric double layer. It was shown that at a sufficiently high surface potential of the electrode, presence of polar or polarizable co-solvent molecules resulted in an increase in the differential capacitance relative to the capacitance in solution without co-solvent impurities, though the polarizability effect appeared to be stronger.

Nevertheless, all the feld-theoretical models of dipolar fluids discussed above treat dipolar particles as point-like ones, i.e. their contribution to the total potential energy of electrostatic interactions is taken into account as a term in the total microscopic charge density described by charge density of point-like dipoles. Such a local dipolar particles model does not allow us to find out how the finite size of the dipolar particles affects the thermodynamic and structural properties of the polar fluids. For instance, the electrostatic potential of the point-like test charge surrounded by point-like dipolar particles in the linear response theory is simply a vacuum Coulomb potential divided by constant dielectric permittivity \cite{Holowko_book}. Thereby, the local theory cannot answer the question: How does the electrostatic potential of point-like test charge depend on the distance at at a scale order of the several dipole lengths? In other words, it is interesting to understand how the electrostatic potential dependence on the distance from the point-like charge deviates from the Coulomb law at the microscopic scales, where the picture of continuous dielectric media becomes invalid.

A very good phenomenological theory named as {\sl nonlocal electrostatics} was formulated by Kornyshev and coauthors \cite{Kornyshev1980}. The central idea of the nonlocal electrostatics approach is the replacement of constant dielectric permeability (or related to it dielectric permittivity) of medium by the nonlocal dielectric permeability (permittivity) operator and rewriting the well known electrostatics equations of the continuous media in its terms. Thus, instead of the well known local Poisson differential equation for the electrostatic potential, the authors obtained a nonlocal integral equation, allowing, in principle (if the functional form of the dielectric permittivity operator is known), one to calculate the electrostatic potential at the microscopic scale. Despite the fact that nonlocal electrostatics allowed the researchers to calculate the electrostatic potential near the charged surfaces of the metals or dielectrics immersed in the electrolyte solutions and the solvation free energy of particles with different electric structures \cite{Kornyshev1980}, it remains unclear how one can calculate within this approach the macroscopic thermodynamic functions of the electrolyte solutions with an explicit account of the dipolar particles.

On the other hand, the microscopic theory of the point-like dipolar particles taking into account the electrostatic correlations at the many-body level (within the Random phase approximation) leads to the ultraviolet divergences of the electrostatic free energy. The latter also indicates the necessity to introduce internal electric structure details of the dipolar particles into the microscopic theory. Up to date, two methods have been proposed for obtaining a finite value of the electrostatic free energy of the dipolar fluid within the microscopic statistical theories. The first approach consists in introducing an ultraviolet cut-off into the integration over the vectors of reciprocal space, so that the cut-off parameter is usually inversely proportional to the molecular length scale (see, for instance \cite{Levy2012,Dean2012,Budkov2016,Budkov2017}). The second approach starts from the assumption that the charge groups of the dipolar particles are not point-like ones described by the delta-functional densities, but are smeared over some spatial region with some form-factor. Thus, introducing the charge form-factor, containing some phenomenological scale parameter to the microscopic theory allows the authors to avoid the ultraviolet divergence \cite{Martin2016}. Note that both approaches give qualitatively similar results for the behavior of electrostatic free energy at a small dipolar particles concentration. Strictly speaking, the first approach is a particular case of the second one, when the Fourier-image of the form-factor choice is the Heaviside step function, which, in turn, leads to the ultraviolet cut-off. We would also like to note that despite the efficiency of these approaches in theoretical description of the thermodynamic properties of different dipolar fluids, they could be considered only as mathematical tricks allowing one to overcome the deficiencies of the local theory without accounting for the internal electric structure of the dipolar particles. However, within both approaches the question stated above about the electrostatic potential behavior at small distances from the test point-like charge remains unanswered.

On the other hand, the recent advances in experimental studies of different organic compounds (such as proteins \cite{Canchi,Haran}, betaines \cite{Kudaibergenov,Lowe}, zwitterionic liquids \cite{Heldebrant2010}, etc.) comprising rather long polar groups ($\sim 5-10~nm$) challenged the statistical mechanics to develop a field-theoretical approach to modelling the dipolar particles not as the point-like dipoles \cite{Budkov2015,Abrashkin2007} (or even as the dipolar hard spheres \cite{BudkovJCP2016,Gongadze2015}), but as a set of charge centers with fixed or fluctuating distances between them. One would expect that such a nonlocal statistical field theory should be devoid of the above described ultraviolet divergence of the free energy and allow us to study the electrostatic potential behavior at the distances from the point-like test charge that are comparable to the characteristic distances between the charge centers of the dipolar particle. In the present paper, we formulate such a nonlocal statistical field theory of the ion-dipole mixture.

\section{Theory}
\subsection{Configuration integral of ion-dipole mixture in functional integral form}
Let us consider an electrolyte solution with $N_{+}$ point-like cations with charge $q_{+}$, $N_{-}$ point-like anions with charge $q_{-}$, and $N_d$ dipolar particles with point-like charged groups with charges $\pm e$ ($e$ is the elementary charge) confined in the volume $V$. Let the numbers of ions $N_{\pm}$ satisfy the global electric neutrality condition $q_{+}N_{+}+q_{-}N_{-}=0$. Let us also assume that the particles are immersed in a solvent which we will model as a uniform dielectric background with constant dielectric permittivity $\varepsilon$. We postulate that each dipolar particle is associated with  the probability distribution function (PDF) $g(\bold{r})$ of the distance between its centers of charge. Keeping in mind these model assumptions, we start from the configuration integral of the mixture
\begin{equation}
Q=\int d\Gamma_d\int d\Gamma_s \exp\left[-\beta H_{el}\right],
\end{equation}
where
\begin{equation}
\int d\Gamma_d(\cdot)=\int\prod\limits_{j=1}^{N_d} \frac{d\bold{r}_j^{(+)}d\bold{r}_j^{(-)}}{V}g(\bold{r}_j^{(+)}-\bold{r}_j^{(-)})(\cdot)
\end{equation}
is the measure of integration over the coordinates $\bold{r}_{j}^{(\pm)}$ of the charge centers of the dipolar particles with the above mentioned PDF $g(\bold{r})$, satisfying the normalization condition
\begin{equation}
\int g(\bold{r}) d\bold{r}=1,
\end{equation}
and
\begin{equation}
\int d\Gamma_s (\cdot)=\frac{1}{V^{N_{+}+N_{-}}}\int\prod\limits_{l=1}^{N_{+}}d\bold{r}_l^{(+)}\int\prod\limits_{k=1}^{N_{-}}d\bold{r}_k^{(-)}(\cdot)
\end{equation}
is the integration measure over the ionic coordinates; $\beta = 1/k_{B}T$ is the inverse thermal energy. The Hamiltonian of the electrostatic interaction can be written as follows
\begin{equation}
\label{hamilt_el}
H_{el}=\frac{1}{2}\int d\bold{r}\int d\bold{r}'\hat\rho(\bold{r})G_0(\bold{r}-\bold{r}')\hat\rho(\bold{r}')=\frac{1}{2}\left(\hat\rho G_0 \hat\rho\right),
\end{equation}
where $G_0(\bold{r}-\bold{r'})=1/(4\pi\varepsilon\varepsilon_0|\bold{r}-\bold{r}'|)$ is the Green function of the Poisson equation
and
\begin{equation}
\hat\rho(\bold r)=e\sum_{j=1}^{N_d}(\delta(\bold{r}-\bold{r}_j^{(+)})-\delta(\bold r-\bold{r}_j^{(-)}))+q_{+}\sum_{k=1}^{N_{+}}\delta(\bold{r}-\bold{r}_k^{(+)})+q_{-}\sum_{l=1}^{N_{-}}\delta(\bold r-\bold{r}_l^{(-)})+\rho_{ext}(\bold r)
\end{equation}
is the total charge density of the system, consisting of the microscopic charge density of the dipolar particles and ions (first, second, and third terms) and the fixed density of external charges (fourth term). It should be noted that we omitted the electrostatic self-energy of the charged particles in (\ref{hamilt_el}) for simplicity. Note that in the present study for simplicity we neglected the contribution of the repulsive interactions between the species. It can be justified for rather small concentrations of salt and dipolar particles.

Thus, we can rewrite our partition function as follows
\begin{equation}
Q=\int d\Gamma_d\int d\Gamma_s \exp\left[-\frac{\beta}{2}(\hat\rho G_0\hat\rho)\right].
\end{equation}
Further, using the Hubbard-Stratonovich transformation (see, for instance, \cite{Podgornik1989,Netz2001})
\begin{equation}
\exp\left[\frac{\beta}{2}(\hat\rho G_0\hat\rho)\right]=\int\frac{\mathcal{D}\varphi}{C}\exp\left[-\frac{\beta}{2}(\varphi G_0^{-1}\varphi)+i\beta(\hat\rho\varphi)\right],
\end{equation}
we achieve the following functional representation of the configuration integral
\begin{equation}
Q=\int\frac{\mathcal{D}\varphi}{C}\exp\left[-\frac{\beta}{2}(\varphi G_0^{-1}\varphi)+i\beta(\rho_{ext}\varphi)\right]Q_d^{N_d}[\varphi]
Q_{+}^{N_{+}}[\varphi]Q_{-}^{N_{-}}[\varphi]
\end{equation}
with the one-particle partition functions
\begin{equation}
Q_d[\varphi]=\int\frac{d\bold r^{(+)}d\bold r^{(-)}}{V}g(\bold r^{(+)}-\bold r^{(-)})\exp(i\beta e(\varphi(\bold r^{(+)})-\varphi(\bold r^{(-)})),
\end{equation}
\begin{equation}
Q_{\pm}[\varphi]=\int\frac{d\bold r}{V}\exp(i\beta q_{\pm}\varphi(\bold r)),
\end{equation}
and the following short-hand notations
$$(\varphi G_0^{-1}\varphi)=\int d\bold r\int d\bold r'\varphi(\bold r)G_0^{-1}(\bold r,\bold r')\varphi(\bold r'),~~(\hat\rho\varphi)=\int d\bold r \hat\rho(\bold r)\varphi(\bold r),$$
$$C=\int \mathcal{D}\varphi\exp\left[-\frac{\beta}{2}(\varphi G_0^{-1}\varphi)\right].$$
In the thermodynamic limit
$$V\to \infty,~N_{d,\pm}\to\infty, N_{d,\pm}/V\to n_{d,\pm}$$
we have \cite{Efimov1996}
\begin{equation}
\nonumber
Q_d^{N_d}[\varphi]=\left[1+\frac{1}{V}\int d\bold r^{(+)}\int d\bold r^{(-)}g(\bold r^{(+)}-\bold r^{(-)})(\exp\left(i\beta e(\varphi(\bold r^{(+)})-\varphi(\bold  r^{(-)}))\right)-1)\right]^{N_d}\simeq
\end{equation}
\begin{equation}
\exp\left[n_d\int d\bold r^{(+)}\int d\bold r^{(-)}g(\bold r^{(+)}-\bold r^{(-)})(\exp\left(i\beta e(\varphi(\bold r^{(+)})-\varphi(\bold  r^{(-)}))\right)-1)\right].
\end{equation}
Similarly, we obtain
\begin{equation}
Q_{\pm}^{N_{\pm}}[\varphi]\simeq\exp\left[n_{\pm}\int d\bold r(\exp(i\beta q_{\pm} \varphi(\bold r))-1)\right].
\end{equation}
Therefore, we arrive at the following functional representation of configuration integral
\begin{equation}
\label{func_int}
Q=\int\frac{\mathcal{D}\varphi}{C}\exp\left[-S[\varphi]\right],
\end{equation}
where the following functional
\begin{equation}
\nonumber
S[\varphi]=\frac{\beta}{2}(\varphi G_0^{-1}\varphi)-i\beta(\rho_{ext}\varphi)-n_{d}\int d\bold r\int d\bold r'g(\bold r-\bold r')(\exp\left[i\beta e(\varphi(\bold r)-\varphi(\bold r'))\right]-1)
\end{equation}
\begin{equation}
\label{functional}
-n_{+}\int d\bold r (\exp\left[i\beta q_{+}\varphi(\bold r)\right]-1)-n_{-}\int d\bold r (\exp\left[i\beta q_{-}\varphi(\bold r)\right]-1)
\end{equation}
is introduced.

Now let us relate the present theory to the above-mentioned Poisson-Boltzmann-Langevin (PBL) theory formulated within the Grand canonical ensemble in references \cite{Abrashkin2007,Coalson1996} and within the Canonical ensemble in reference \cite{Budkov2015}. Determining the PDF for the dipolar particles with a fixed distance between the charge centers $|\bold{r}-\bold{r}'|=l$ in the following form
\begin{equation}
\label{hard_diploes}
g(\bold r-\bold r')=\int\frac{d\bold{n}}{4\pi}\delta(\bold r-\bold r'-\bold l),
\end{equation}
where integration $\int d\bold{n}/4\pi(..)$ means the averaging over orientations of unit vector $\bold{n}=\bold{l}/l$, we obtain the following functional
\begin{equation}
\nonumber
S[\varphi]=\frac{\beta}{2}(\varphi G_0^{-1}\varphi)-n_d\int\frac{d\bold{n}}{4\pi}\int d\bold r (\exp\left(i\beta e(\varphi(\bold r)-\varphi(\bold r-\bold l))\right)-1)
\end{equation}
\begin{equation}
-n_{+}\int d\bold r (\exp\left[i\beta q_{+}\varphi(\bold r)\right]-1)-n_{-}\int d\bold r (\exp\left[i\beta q_{-}\varphi(\bold r)\right]-1)-i\beta(\rho_{ext}\varphi)
\end{equation}

When $\bold l \to 0$, the expansion of the exponential factor into the Taylor series up to the first order in $\bold{l}$ and averaging over orientations of vector $\bold{l}$ lead to the well known PBL functional \cite{Abrashkin2007,Coalson1996,Budkov2015}
\begin{equation}
\nonumber
S_{PBL}[\varphi]=\frac{\beta}{2}(\varphi G_0^{-1}\varphi)-n_d\int d\bold r\left(\frac{\sin \beta p|\bold \nabla\varphi(\bold r)|}{\beta p|\bold \nabla\varphi(\bold r)|}-1\right)
\end{equation}
\begin{equation}
-n_{+}\int d\bold r (\exp\left[i\beta q_{+}\varphi(\bold r)\right]-1)-n_{-}\int d\bold r (\exp\left[i\beta q_{-}\varphi(\bold r)\right]-1)-i\beta(\rho_{ext}\varphi),
\end{equation}
where $p=el$ is the dipole moment.

\subsection{Mean-field approximation}
Now let us consider the mean-field approximation for the configuration integral \cite{Podgornik1989,Lue2006,Budkov2015}. At first, we have to equate to zero the functional derivative of the functional (\ref{functional}), i.e.
\begin{equation}
\frac{\delta S[\varphi]}{\delta\varphi(\bold r)}=0,
\end{equation}
that yields the following equation
\begin{equation}
\nonumber
\int d\bold r'G_0^{-1}(\bold r-\bold r')\varphi(\bold r')-i\rho_{ext}(\bold r)+2n_d e\int d\bold r' g(\bold r-\bold r')\sin\frac{e(\varphi(\bold r)-\varphi(\bold r'))}{k_BT}
\end{equation}
\begin{equation}
-i\left(q_{+}n_{+}\exp\left[i\beta q_{+}\varphi(\bold r)\right]+q_{-}n_{-}\exp\left[i\beta q_{-}\varphi(\bold r)\right]\right)
=0.
\end{equation}
Further, introducing the electrostatic potential $\varphi(\bold{r})=i\psi(\bold{r})$ \cite{Podgornik1989,Budkov2015,Lue2006}, we arrive at the following nonlinear integro-differential self-consistent field equation:
\begin{equation}
\nonumber
-\varepsilon\varepsilon_0\Delta\psi(\bold r)=2n_d e\int d\bold r'g(\bold r-\bold r')\sinh\frac{e(\psi(\bold r')-\psi(\bold r))}{k_BT}
\end{equation}
\begin{equation}
\label{self-cons_eq}
+q_{+}n_{+}\exp\left[-\beta q_{+}\psi(\bold{r})\right]+q_{-}n_{-}\exp\left[-\beta q_{-}\psi(\bold{r})\right]+\rho_{ext}(\bold r).
\end{equation}
The electrostatic free energy in the mean-field approximation is
\begin{equation}
F_{el}^{(MF)}=k_BTS[i\psi].
\end{equation}

\subsection{Point-like charge in dipole environment}
As an application of equation (\ref{self-cons_eq}), let us calculate the electrostatic potential of the electric field created by a point-like test ion with charge $q$ surrounded by dipolar particles in the absence of ions (i.e., $n_{\pm}=0$). Placing the point-like ion at the origin and taking into account that $\rho_{ext}(\bold{r})=q\delta(\bold{r})$, we write the self-consistent field equation in the form
\begin{equation}
\label{self-cons_eq2}
-\varepsilon\varepsilon_0\Delta\psi(\bold r)=2n_d e\int d\bold r'g(\bold r-\bold r')\sinh\frac{e(\psi(\bold r')-\psi(\bold r))}{k_BT}+q\delta(\bold r).
\end{equation}
In order to understand how the dipolar surrounding changes the Coulomb electrostatic potential of the bare point-like charge, we consider the regime of weak electrostatic interaction, {\sl i.e.} assume that $e\psi(\bold r)/{k_BT}\ll1$, which yields
\begin{equation}
\label{self-cons_eq_lin}
-\varepsilon\varepsilon_0\Delta\psi(\bold r)=-\frac{2n_{d}e^2}{k_BT}\psi(\bold r)+\frac{2n_{d}e^2}{k_BT}\int d\bold r'g(\bold r-\bold r')\psi(\bold r')+q\delta(\bold r).
\end{equation}
Further, passing to the Fourier-image of the electrostatic potential
$$\tilde{\psi}(\bold k)=\int d\bold r\psi(\bold r)e^{-i\bold k\bold r},$$
we can rewrite the equation (\ref{self-cons_eq_lin}) in the following algebraic form
\begin{equation}
\varepsilon\varepsilon_0k^2\tilde\psi(\bold k)=-\frac{2n_{d}e^2}{k_BT}(1-g(\bold k))\tilde\psi(\bold k)+q,
\end{equation}
where $k=|\bold{k}|$. Thus, after some algebraic transformations, we arrive at
\begin{equation}
\tilde\psi(\bold k)=\frac{q}{\varepsilon\varepsilon_0(k^2+\varkappa^2(\bold k))},
\end{equation}
where
\begin{equation}
\label{screen_func}
\varkappa^2(\bold k)=\frac{2n_{d}e^2}{\varepsilon\varepsilon_0k_BT}(1-g(\bold k))
\end{equation}
is the so-called screening function and
\begin{equation}
g(\bold k)=\int d\bold{r}e^{-i\bold{k}\bold{r}}g(\bold{r})
\end{equation}
is the characteristic function of distribution \cite{Gnedenko}.
Therefore, the electrostatic potential of the point charge takes the standard form of the linear response theory \cite{Barrat_Hansen,Khokhlov1982,Borue1988}:
\begin{equation}
\label{pot_lin}
\psi(\bold r)=\frac{q}{\varepsilon\varepsilon_0}\int\frac{d\bold k}{(2\pi)^3}\frac{e^{i\bold k\bold r}}{k^2+\varkappa^2(\bold k)}.
\end{equation}
In order to evaluate the electrostatic potential, we should specify the probability distribution function $g(\bold{r})$. For the dipolar particles with permanent dipole moments with the PDF determined by relation (\ref{hard_diploes}) the characteristic function takes the form
\begin{equation}
\label{hard_dip_char}
g(\bold{k})=\frac{\sin{kl}}{kl}.
\end{equation}
However, relation (\ref{hard_dip_char}) does not allow us to obtain the analytical expression for electrostatic potential. Nevertheless, we can use another model characteristic function
\begin{equation}
\label{char_func_model}
g(\bold k)=\frac{1}{1+\frac{k^2l^2}{6}},
\end{equation}
which allows us to obtain the analytic results, determining the following 'Yukawa'-type PDF
\begin{equation}
\label{soft_dip_char}
g(\bold r)=\frac{3}{2\pi l^2 r}\exp\left(-\frac{\sqrt{6}r}{l}\right),
\end{equation}
where $r=|\bold{r}|$. We would like to stress that the distributions determined by characteristic functions (\ref{hard_dip_char}) and (\ref{char_func_model}) give the same mean-square distances between the charge centers of dipolar particles $l/\sqrt{6}$.

Thus, substituting eq. (\ref{char_func_model}) with eq. (\ref{screen_func}) and taking integral (\ref{pot_lin}), we arrive at the following result
\begin{equation}
\label{potential_fin}
\psi(\bold{r})=\frac{q}{4\pi\varepsilon\varepsilon_0r}\frac{1+y_d\exp\left(-\frac{r}{l_s}\right)}{1+y_d},
\end{equation}
where $y_d=n_{d}e^2l^2/(3\varepsilon\varepsilon_0k_BT)=l^2/6 r_{D}^2$ is the electrostatic $"$strength$"$ of dipole-dipole electrostatic interaction and $l_s=l/\sqrt{6\varepsilon(1+y_d)}$ is the length, whose physical sense will be clarified below; $r_{D}=\left(2n_d e^2/\varepsilon\varepsilon_{0}k_{B}T\right)^{-1/2}$ is the effective Debye screening length attributed to the charge centers of dipolar particles. It is interesting to investigate the electrostatic potential behavior for very long dipolar particles, i.e when $y_d \gg 1$ (or $l\gg r_D$). In that regime, relation (\ref{potential_fin}) takes the form of the well known Debye-Hueckel potential
\begin{equation}
\psi(\bold{r})\simeq \frac{q}{4\pi\varepsilon\varepsilon_0r}\exp\left(-\frac{r}{r_D}\right).
\end{equation}
Thus, when the mean-square dipole length is much bigger than the Debye screening length, the charge centers of dipolar particles can be considered as unbonded ions, screening the charge of the test ion.

It is worth noting that relation (\ref{potential_fin}) allows us to formally introduce the local dielectric permittivity as
\begin{equation}
\varepsilon(\bold{r})=\frac{\varepsilon(1+y_d)}{1+y_d\exp\left(-\frac{r}{l_s}\right)}.
\end{equation}
Therefore, length $l_s$ determines the radius of the sphere around the point charge in which the dielectric constant is smaller than its bulk value \cite{Ramshaw1978,Ramshaw1980,Abrashkin2007,Budkov2015}
\begin{equation}
\label{eps_bulk}
\varepsilon_b=\varepsilon(1+y_d)=\varepsilon+\frac{n_d e^2l^2}{3\varepsilon_0k_BT}.
\end{equation}
In other words, length $l_{s}$ can be considered as the effective solvation radius of the point charge in the dipole environment within the linear theory.

The mean-field electrostatic free energy in the linear approximation is:
\begin{equation}
\nonumber
F_{el}^{(MF)}=k_BTS[i\psi]=-\frac{1}{2}\int d\bold r\int d\bold r'\psi(\bold r)G_0^{-1}(\bold r,\bold r')\psi(\bold r')
\end{equation}
\begin{equation}
\nonumber
-\frac{n_d e^2}{2k_BT}\int d\bold r\int d\bold r'g(\bold r-\bold r')(\psi(\bold r)-\psi(\bold r'))^2+\int d\bold r\rho_{ext}(\bold r)\psi(\bold r)=
\end{equation}\
\begin{equation}
-\frac{1}{2}\int d \bold r\int d\bold r'\psi(\bold r)G_{d}^{-1}(\bold r,\bold r')\psi(\bold r')+\int d \bold r\rho_{ext}(\bold r)\psi(\bold r),
\end{equation}
where the reciprocal Green function renormalized by the presence of dipolar particles
\begin{equation}
G_{d}^{-1}(\bold r,\bold r')=G_0^{-1}(\bold r,\bold r')+\frac{2n_d e^2}{k_BT}(\delta(\bold r-\bold r')-g(\bold r-\bold r'))
\end{equation}
is introduced.
Further, taking into account that
$$\rho_{ext}(\bold r)=\int d\bold r' G_{d}^{-1}(\bold r,\bold r')\psi(\bold r'),$$
we obtain the final relation:
\begin{equation}
\label{el_free_en_lin}
F_{el}=\frac{1}{2}\int d\bold r\rho_{ext}(\bold r)\psi(\bold r).
\end{equation}
To calculate the solvation free energy of the point-like test ion, we should subtract the electrostatic self-energy of point-like charge, i.e.
\begin{equation}
\label{solv}
\Delta F_{solv}=\frac{1}{2}\int d\bold r\rho_{ext}(\bold r)(\psi(\bold r)-\psi_{ext}(\bold r))
\end{equation}
from (\ref{el_free_en_lin}).
Taking into account that
\begin{equation}
\label{psi_ext}
\psi_{ext}(\bold r)=\frac{q}{4\pi\varepsilon\varepsilon_0r},
\end{equation}
we arrive at
\begin{equation}
\label{solv_2}
\Delta F_{solv}=-\frac{q^2\sqrt6}{8\pi\varepsilon\varepsilon_0l}\frac{y_d}{\sqrt{1+y_d}}.
\end{equation}

\subsection{Point-like charge in ion-dipole environment}
Now let us calculate the electrostatic potential of the point-like charge at the non-zero ionic concentrations in the bulk ($n_{\pm}\neq 0$) in the linear approximation $e\psi/k_{B}T\ll 1$. In that case the self-consistent field equation takes the following linearized form
\begin{equation}
-\varepsilon\varepsilon_{0}\Delta \psi(\bold{r})=-\frac{2 I e^2 }{k_BT}\psi(\bold{r})-\frac{2n_{d}e^2}{k_BT}\psi(\bold r)+\frac{2n_{d}e^2}{k_BT}\int d\bold r'g(\bold r-\bold r')\psi(\bold r')+q\delta(\bold r),
\end{equation}
where $I=\left(q_{+}^2 n_{+}+q_{-}^2 n_{-}\right)/2e^2$ is the solution ionic strength.

Moving on to the Fourier representation, after some algebra, we obtain the standard relation for the Fourier-image of the potential
\begin{equation}
\label{pot_lin_salt}
\tilde{\psi}(\bold{k})=\frac{q}{\varepsilon\varepsilon_{0}\left(k^2+\varkappa^2(\bold{k})\right)}
\end{equation}
with the following screening function
\begin{equation}
\label{screen_func_2}
\varkappa^2(\bold{k})=\kappa_{s}^2+\frac{2n_{d}e^2}{\varepsilon\varepsilon_{0}k_{B}T}\left(1-g(\bold{k})\right),
\end{equation}
where $\kappa_{s}=\sqrt{2e^2I/(\varepsilon\varepsilon_{0}k_{B}T)}$ is the inverse Debye screening length attributed to the ionic species.

Using the same Yukawa-type PDF, i.e., using the same characteristic function $g(\bold{k})=1/\left(1+k^2l^2/6\right)$ as in the previous subsection and performing the inverse Fourier transformation, we arrive at the following analytic relation for the electrostatic potential
\begin{equation}
\label{potential_fin_2}
\psi(\bold{r})=\frac{q}{4\pi\varepsilon\varepsilon_{0} r}
\left(u(y_{d},y_{s})e^{-\kappa_1(y_{d},y_{s})r}+\left(1-u(y_{d},y_{s})\right)e^{-\kappa_2(y_{d},y_{s})r}\right),
\end{equation}
where
\begin{equation}
\kappa_{1,2}(y_{d},y_{s})=\frac{\sqrt{3}}{l}\left(1+y_s+y_d \pm \sqrt{(1+y_s+y_d)^2-4y_s}\right)^{1/2}, ~ y_{s}=\frac{\kappa_{s}^2l^2}{6},
\end{equation}
\begin{equation}
u(y_{d},y_{s})=\frac{y_s+y_d+\sqrt{(1+y_s+y_d)^2-4y_s}-1}{2\sqrt{(1+y_s+y_d)^2-4y_s}}.
\end{equation}
In the absence of ions ($y_{s}=0$), expression (\ref{potential_fin_2}) will transform into the above expression (\ref{potential_fin}). In the absence of dipolar particles ($y_{d}=0$), we obtain the standard Debye-Hueckel potential
\begin{equation}
\psi(\bold{r})=\frac{q}{4\pi\varepsilon\varepsilon_{0}r}e^{-\kappa_{s}r}.
\end{equation}

In order to obtain the solvation free energy of the test charge in the ion-dipole environment, we execute the similar algebraic transformations as in the previous subsection. In that case, we have the same relation for the electrostatic free energy in the mean-field approximation, i.e.
\begin{equation}
\label{el_free_en_lin_1}
F_{el}^{(MF)}=-\frac{1}{2}\int d \bold r\int d\bold r'\psi(\bold r)G^{-1}(\bold r,\bold r')\psi(\bold r')+\int d \bold r\rho_{ext}(\bold r)\psi(\bold r)=\frac{1}{2}\int d\bold r\rho_{ext}(\bold r)\psi(\bold r),
\end{equation}
where
\begin{equation}
G^{-1}(\bold r,\bold r')=G_0^{-1}(\bold r,\bold r')+\frac{2I e^2}{k_BT}\delta(\bold r-\bold r')+\frac{2n_d e^2}{k_BT}(\delta(\bold r-\bold r')-g(\bold r-\bold r'))
\end{equation}
is the reciprocal Green function renormalized by the presence of ions and dipolar particles.

Subtracting the electrostatic self-energy from (\ref{el_free_en_lin_1}) (see the previous subsection), we obtain the following analytic expression
\begin{equation}
\Delta F_{solv}=-\frac{q^2}{8\pi\varepsilon\varepsilon_{0}}\left(u(y_{d},y_{s})(\kappa_1(y_{d},y_{s})-\kappa_2(y_{d},y_{s}))+\kappa_2(y_{d},y_{s})\right),
\end{equation}
which will turn into the above relation (\ref{solv_2}) in the ion-free case ($y_s=0$).

\subsection{Random phase approximation}
Now we proceed to the derivation of the electrostatic free energy of the ion-dipole mixture within the Random phase approximation (RPA). Expanding the functional $S[\varphi]$ in (\ref{func_int}) into a power series near the mean-field $\varphi(\bold{r})=i\psi(\bold{r})$ and truncating it at the quadratic term, we obtain
\begin{equation}
S[\varphi]\approx S[i\psi]+\frac{\beta}{2}\left(\varphi\mathcal{G}^{-1}\varphi\right),
\end{equation}
where
\begin{equation}
\nonumber
\mathcal{G}^{-1}(\bold r,\bold r'|\psi)=k_{B}T\frac{\delta^2 S[i\psi]}{\delta \varphi(\bold{r})\delta \varphi(\bold{r}')}
\end{equation}
\begin{equation}
\nonumber
=G_0^{-1}(\bold r,\bold r')+\frac{1}{k_BT}\left(q_{+}^2n_{+}\exp\left(-\beta q_{+}\psi(\bold{r})\right)+q_{-}^2n_{-}\exp\left(-\beta q_{-}\psi(\bold{r})\right)\right)\delta(\bold r-\bold r')
\end{equation}
\begin{equation}
+\frac{2n_d e^2}{k_BT}\int d\bold{r}''g(\bold{r}-\bold{r}'')\cosh\left(\frac{e\left(\psi(\bold{r})-\psi(\bold{r}'')\right)}{k_{B}T}\right)(\delta(\bold r-\bold r')-\delta(\bold r''-\bold r'))
\end{equation}
is the renormalized reciprocal Green function with the mean-field electrostatic potential $\psi(\bold{r})$ satisfying equation (\ref{self-cons_eq}).
Therefore, taking the Gaussian functional integral \cite{Podgornik1989}, we obtain the following general relation for the configuration integral in the RPA
\begin{equation}
\label{RPA_gen}
Q\approx \exp\left\{-S[i\psi]+\frac{1}{2}tr\left(\ln \mathcal{G} -\ln G_{0}\right)\right\},
\end{equation}
where the symbol $tr(..)$ means the trace of the integral operator \cite{Podgornik1989,Lue2006}.
In the absence of external charges ($\rho_{ext}(\bold{r})=0$), the mean-field electrostatic potential $\psi(\bold{r})=0$, so that the mean-field contribution to the electrostatic free energy $F_{el}^{(MF)}=k_{B}T S[0]=0$. Therefore, the electrostatic free energy is fully determined by the electrostatic potential fluctuations near its zero value. In that case the renormalized reciprocal Green function takes the form
\begin{equation}
\mathcal{G}^{-1}(\bold r,\bold r'|0)=G^{-1}(\bold r,\bold r')=G_0^{-1}(\bold r,\bold r')+\frac{2I e^2}{k_BT}\delta(\bold r-\bold r')+\frac{2n_d e^2}{k_BT}(\delta(\bold r-\bold r')-g(\bold r-\bold r')),
\end{equation}
whereas the electrostatic free energy is \cite{Podgornik1989}
\begin{equation}
F_{el}\approx \frac{k_{B}T}{2}tr\left(\ln G_0 - \ln G\right)=\frac{Vk_{B}T}{2}\int\frac{d\bold k}{(2\pi)^3}\ln\frac{G_0(\bold k)}{G(\bold k)},
\end{equation}
where $G_0(\bold k)=1/(\varepsilon\varepsilon_0k^2)$ and $G(\bold k)=1/(\varepsilon\varepsilon_0(k^2+\varkappa^2(\bold k)))$ are the Fourier-images of the Green functions. Hence, the electrostatic free energy of the ion-dipole mixture in the RPA can be written in the standard form \cite{Borue1988,Lue2006}
\begin{equation}
\label{el_free_en_lin_2}
F_{el}=\frac{Vk_BT}{2}\int\frac{d\bold k}{(2\pi)^3}\left(\ln\left(1+\frac{\varkappa^2(\bold k)}{k^2}\right)-\frac{\varkappa^2(\bold k)}{k^2}\right),
\end{equation}
where the screening function $\varkappa^2(\bold{k})$ is determined by relation (\ref{screen_func_2}). Note that we have subtracted the electrostatic self-energy of the system from the electrostatic free energy \cite{Borue1988}. We would like to stress that integral (\ref{el_free_en_lin_2}) converges at the ultraviolet limit.

For the Yukawa-type PDF, determined by eq. (\ref{char_func_model}), the integral in (\ref{el_free_en_lin_2}) can be taken analytically only in the absence of ions in the system ($n_{\pm}=0$), yielding
\begin{equation}
\label{RPA_pure_dip}
F_{el}=-\frac{Vk_BT}{l^3}\sigma(y_{d}),
\end{equation}
where the auxiliary dimensionless function
\begin{equation}
\label{sigma}
\sigma(y_d)=\frac{\sqrt6}{4\pi}(2(1+y_{d})^{3/2}-2-3y_{d})
\end{equation}
is introduced. It is instructive to analyze the limiting regimes following from relations (\ref{RPA_pure_dip}-\ref{sigma}). Thus, we get
\begin{equation}
\label{analysis_RPA}
\frac{F_{el}}{Vk_BT}=
\begin{cases}
-\frac{\sqrt{6}e^4 l n_{d}^2}{48\pi \left(\varepsilon\varepsilon_{0}k_{B}T\right)^2}, y_d\ll 1\,\\
-\frac{1}{12\pi r_{D}^3},y_d\gg 1.
\end{cases}
\end{equation}
The first regime determines the case when the mean-square distance $l/\sqrt{6}$ is much less than the Debye screening length $r_{D}=\left(2n_d e^2/\varepsilon\varepsilon_{0}k_{B}T\right)^{-1/2}$ introduced above and attributed to the charged groups of dipolar particles. In that case, the electrostatic correlations consist of the pairwise effective Keesom interaction of the dipolar particles \cite{Keesom1916,Reinganum1912}. In the opposite case, when the mean-square distance $l/\sqrt{6}$ is much bigger than the Debye screening length $r_D$, the electrostatic free energy can be described by the standard Debye-Hueckel (DH) limiting law. The latter strictly justifies the idea proposed first in work \cite{Levin1999} that charged groups of very long dipolar particles can be modelled as unbonded point-like ions.

Now let us consider the limiting behavior of the electrostatic free energy in the presence of salt at $y_{d}\ll 1$ and $y_{d}\gg 1$. At $y_{d}\ll 1$ we have
\begin{equation}
\frac{F_{el}}{Vk_{B}T}=-\frac{\kappa_{s}^3}{12\pi}-\frac{3\sqrt{6}}{2\pi l^3}\frac{y_{s}}{1+\sqrt{y_s}}y_d-\frac{3\sqrt{6}}{16\pi l^3}\frac{1+3\sqrt{y_s}+y_s}{(1+\sqrt{y_s})^3}y_{d}^2+O(y_{d}^3),
\end{equation}
where the first term in the right hand side describes the DH contribution to the electrostatic free energy from the ionic component of the mixture. The second and third terms describe, respectively, the contributions of ion-dipole and dipole-dipole pairwise correlations. In the opposite regime when $y_d\gg 1$, we obtain the standard DH limiting law
\begin{equation}
\frac{F_{el}}{Vk_{B}T}\simeq-\frac{1}{12\pi}\left(\frac{2(n_d +I)e^2}{\varepsilon\varepsilon_{0}k_{B}T}\right)^{3/2}.
\end{equation}
In the latter case, dipolar particles charged groups manifest themself as unbonded ions participating in the charge screening together with salt ions.

\section{Concluding remarks and perspectives}
In conclusion, we have formulated a nonlocal statistical field theory of the dilute electrolyte solutions with admixture of a small amount of dipolar particles, consideringmodelling the latter as dimers comprising two oppositely charged groups located at the fluctuating distances relative from each other. Attributing to every dipolar particle an arbitrary probability distribution function of the distance between its charge centers and using the standard Hubbard-Stratonovich transformation, we have represented the configuration integral of the system in the functional integral form with the functional in integrand exponent, generalizing the Poisson-Boltzmann-Langevin functional \cite{Abrashkin2007,Coalson1996,Budkov2015} obtained earlier for the ions mixed with point-like dipolar particles. Within the mean-field approximation we have derived a nonlinear integro-differential equation for the mean-field potential. Linearizing the obtained self-consistent field equation, we have derived the general expressions for the electrostatic potential around the point-like test ion and its solvation free energy in the solution of salt-free dipolar particles, as well as in the solution with the nonzero salt concentration. For the 'Yukawa'-type probability distribution function we have obtained analytic expressions for the electrostatic potential of the point-like test ion and its solvation free energy. We have obtained a general relation for the bulk electrostatic free energy in the Random phase approximation expressed through the integral over the wave vectors of the reciprocal space converging at the ultraviolet limit. For Yukawa-type probability distribution function we have obtained an analytic relation for the electrostatic free energy of the salt-free solution of dipolar particles, analyzing all limiting regimes that follow from it. We have also analyzed the limiting regimes, following from the general relation for the electrostatic free energy of the dipolar particles solution in case of the Yukawa-type probability distribution function in the presence of salt.

Finally, we would like to discuss further applications of the present theory. As is well known, in the dipolar fluids at a rather low temperature chalin-like clusters are formed, preventing the expected liquid-gas phase separation \cite{Weis1993,Levin1999,Levin2001}. In contrast to the first, quite eclectic theory \cite{Levin1999,Levin2001} taking into account the chain-like cluster formation in the dipolar fluids, the present theory allows us, in principle, to calculate the electrostatic free energy of the dipolar chain clusters mixture within one formalism. Thus, it is interesting to apply the present formalism to the description of dipolar fluids phase behavior taking into account for the cluster formation of the dipolar particles and their dispersion and excluded volume interactions. Especially, it is interesting to answer the following question: How will accounting for the chain-like cluster formation change the self-consistent field equation (\ref{self-cons_eq})? Answering that question will give us an understanding of how the presence of chain-like dipolar clusters in a polar fluid will renormalize the electrostatic potential of the test ion and its solvation free energy. There are also two advantages of the present theory over the previously published statistical theory of polar fluids \cite{Levin1999,Levin2001}. Firstly, in contrast to that theoretical model, the present theory can be easily generalized to the case of dipolar particles mixture with an arbitrary number of components. Secondly, our theory can be generalized to the particles with an arbitrary electric structure. For this purpose, it is necessary to introduce a probability distribution function of distance for each pair of oppositely charged groups of the fluid molecule. A systematic study of all these issues will be the subject of forthcoming publications.

\begin{acknowledgements}
The author has benefited from conversations with I.Y. Erukhimovich and A.A. Kornyshev.
\end{acknowledgements}


\newpage
\begin{thebibliography}{00}
\bibitem{Onsager1936}
Onsager L 1936 {\it J. Am. Chem. Soc.} {\bf 58} (8) 1486

\bibitem{Kirkwood1939}
Kirkwood J 1939 {\it J. Chem. Phys.} {\bf 7} 911

\bibitem{Hoye1976}
Hoye J S and Stell G 1976 {\it J. Chem. Phys.} {\bf 64} 1952

\bibitem{Chandler1977}
Chandler D 1977 {\it J. Chem. Phys.} {\bf 67} 1113

\bibitem{Holowko_book}
Iukhnovskii I R, Golovko M F 1980 {\it Statistical theory of classical equilibrium systems} (Kiev, Izdatel'stvo Naukova Dumka, In Russian)

\bibitem{Levin1999}
Levin Y 1999 {\it PRL} {\bf 83} (6) 1159

\bibitem{Fisher1993}
Fisher M, Levin Y 1993 {\it PRL} {\bf 71} 3826

\bibitem{Nienhuis1972}
Nienhuis G and Deutch J M 1972 {\it J. Chem. Phys.} {\bf 56} 235

\bibitem{Weiss1998}
Weiss V C and  Schroer W 1998 {\it J. Chem. Phys.} {\bf 108} 7747

\bibitem{Abrashkin2007}
Abrashkin A, Andelman D and Orland H 2007 {\it PRL} {\bf 99} 077801

\bibitem{Coalson1996}
Coalson R D, Duncan A 1996 {\it J. Phys. Chem.} 1996 {\bf 100} 2612

\bibitem{Budkov2015}
Budkov Yu A, Kolesnikov A L and Kiselev M G 2015 {\it EPL} {\bf 111} 28002


\bibitem{Kornyshev1980}
Kornyshev A A and Vorotyntsev M A 1980 {\it Surface Science} {\bf 101} 23

\bibitem{Levy2012}
Levy A, Andelman D and Orland H 2012 {\it PRL} {\bf 108} 227801

\bibitem{Dean2012}
Dean D S and Podgornik R 2012 {\it J. Chem. Phys.} {\bf 136} 154905

\bibitem{Budkov2016}
Budkov Yu A and Kolesnikov A L 2016 {\it Eur. Phys. J. E} {\bf 39} 110

\bibitem{Budkov2017}
Budkov Yu A, Kalikin N N and Kolesnikov A L 2017 {\it Eur. Phys. J. E}  {\bf 40} 47

\bibitem{Martin2016}
Martin J M, Li W, Delaney K T, and Fredrickson G H 2016 {\it J. Chem. Phys.} {\bf 145} 154104

\bibitem{Canchi}
Canchi D R and Garcia A E 2013 {\it Annu. Rev. Phys. Chem.} {\bf 64} 273

\bibitem{Haran}
Haran G 2012 {\it Curr. Opin. Struct. Biol.} {\bf 22} 14

\bibitem{Lowe}
Lowe A B and McCormick C L 2002 {\it Chem. Rev.} {\bf 102} 4177

\bibitem{Kudaibergenov}
Kudaibergenov S, et al 2006 {\it Advances in Polymer Science} {\bf 201} 157

\bibitem{Heldebrant2010}
Heldebrant D J, et al 2010 {\it Green Chemistry} {\bf 12} 713

\bibitem{BudkovJCP2016}
Budkov Y A, Kolesnikov A L, Kiselev M G 2016 {\it J. Chem. Phys.} 144 184703

\bibitem{Gongadze2015}
Gongadze E, Iglic A. 2015 {\it Electrochim. Acta} 178 541

\bibitem{Podgornik1989}
Podgornik R 1989 {\it J. Chem. Phys.} {\bf 91} 5840

\bibitem{Lue2006}
Lue L 2006 {\it Fluid Phase Equilibria} {\bf 241} 236

\bibitem{Netz2001}
Netz R R 2001 {\it Eur. Phys. J. E} {\bf 5} 557

\bibitem{Efimov1996}
Efimov G V, Nogovitsin E A 1996 {\it Physica A} {\bf 234} 506

\bibitem{Gnedenko}
Gnedenko B V 2001 {\it The theory of probability and the elements of statistics} (Fifth edition, AMS Chelsea Publishing)

\bibitem{Barrat_Hansen}
Barrat J L and Hansen J P 2003 {\it Basic Concepts for Simple and Complex Liquids} (Cambridge: University Press)

\bibitem{Khokhlov1982}
Khokhlov A R, Khachaturian K A 1982 {\it Polymer} {\bf 23} (12) 1742

\bibitem{Borue1988}
Borue V Yu and Erukhimovich I Ya 1988 {\it Macromolecules} {\bf 21} 3240


\bibitem{Ramshaw1978}
Ramshaw J D 1978 {\it J. Chem. Phys.} {\bf 68} 5199

\bibitem{Ramshaw1980}
Ramshaw J D 1980 {\it J. Chem. Phys.} {\bf 73} 3695

\bibitem{Reinganum1912}
Reinganum M 1912 {\it Ann. Phys.} (Leipzig) 38 649

\bibitem{Keesom1916}
Keesom W H 1916 {\it Proc. K. Ned. Akad. Wet.} {\bf 18} 636

\bibitem{Weis1993}
Weis J J and Levesque D 1993 {\it PRL} {\bf 71} 2729

\bibitem{Levin2001}
Levin Y, Kuhna P S, Barbosa M C 2001 {\it Physica A} {\bf 292} 129

\end{thebibliography}
\end{document}